\documentstyle{article}
\def\be{\begin{equation}}
\def\ee{\end{equation}}
\def\bea{\begin{eqnarray}}
\def\eea{\end{eqnarray}}

\def\beq{\begin{equation}}
\def\eeq{\end{equation}}
\def\bea{\begin{eqnarray}}
\def\eea{\end{eqnarray}}
\def\nn{\nonumber}
\def\ba{\begin{array}}
\def\ea{\end{array}}   
\def\0{{\mbox{\boldmath $0$}}}

\def\A{{\mbox{\boldmath $A$}}}
\def\B{{\mbox{\boldmath $B$}}}

\def\S{{\mbox{\boldmath $S$}}}

\def\vpi{{\mbox{\boldmath $\pi$}}}
\def\r{{\mbox{\boldmath $r$}}}

\def\i{{\rm i}}

\def\pitwo{\hat{\pi}_\perp^{\,2}}

\def\ddz{\frac{\partial}{\partial z}} 
\def\vsig{{\mbox {\boldmath $\sigma$ }}} 
\def\Al{{\mbox{\boldmath $\alpha$}}}

\def\half{\frac{1}{2}}

\def\loh{\lambda_0}

\def\eps{\epsilon}
\def\g{\gamma}
\def\Vomeg{{\underline{\mbox {\boldmath $\Omega$}}}_s}

\def\hH{\hat{H}} 
\def\Vsig{{\mbox {\boldmath $\Sigma$ }}}

\begin{document}
\begin{flushright}
physics/9904063
\end{flushright}

\begin{center}

{\large\bf Quantum aspects of accelerator optics}
\footnote{In Proceedings of the 1999 Particle Accelerator Conference~(PAC99)
29 March -- 02 April 1999, New York City,
Editors: A.~Luccio and W.~MacKay}

\bigskip

Sameen Ahmed KHAN \\

Dipartimento di Fisica Galileo Galilei  
Universit\`{a} di Padova \\
Istituto Nazionale di Fisica Nucleare~(INFN) Sezione di Padova \\
Via Marzolo 8 Padova 35131 ITALY \\
E-mail: khan@pd.infn.it, ~~~ http://www.pd.infn.it/$\sim$khan/

\end{center}

\bigskip

\begin{abstract}
Present understanding of accelerator optics is based mainly on classical 
mechanics and electrodynamics. In recent years  quantum theory of 
charged-particle beam optics has been under development. In this paper 
the newly developed formalism is outlined.
\end{abstract}

\bigskip


\bigskip


\newpage

Charged-particle beam optics, or the theory of transport of charged-particle
beams  through electromagnetic systems, is traditionally dealt with using 
classical mechanics. This is the case in ion optics, electron microscopy,
accelerator physics etc~\cite{HK}-\cite{Wiedemann}. The classical
treatment of charged-particle beam optics has been extremely successful, in
the designing and working of numerous optical devices from electron microscopes
to very large particle accelerators, including polarized beam accelerators. 
It is natural to look for a prescription
based on the quantum theory, since any physical system is quantum at the
fundamental level! Such a prescription is sure to explain the grand success of 
the classical theories and may also help towards a deeper understanding and 
designing of certain charged-particle beam devices. To date the curiosity to
justify the success of the classical theories as a limit of a quantum theory 
has been the main motivation to look for a quantum prescription. But, with ever
increasing demand for higher luminosities and the need for polarized beam
accelerators in basic physics, we strongly believe that the quantum theories,
which up till now were an isolated academic curiosity will have a significant
role to play in designing and working of such devices.

It is historically very curious that the, quantum approaches to the
charged-particle beam optics have been very modest and have a very brief
history as pointed out in the third volume of the three-volume encyclopaedic
text book of Hawkes and Kasper~\cite{HK3}. In the context of accelerator 
physics the grand success of the classical theories originates from the fact
that {\em the de Broglie wavelength of the} ({\em high energy}) 
{\em beam particle is very small compared to the typical apertures of
the cavities in accelerators}. 
This and related details have been pointed out 
in the recent article of Chen~\cite{Chen}. 
A detailed account of the {\bf quantum aspects of beam physics} is to be
found in the Proceedings of the recently held 15th Advanced ICFA Beam 
Dynamics Workshop~\cite{QABP}.

A beginning of a quantum formalism starting {\em ab initio} with the 
Dirac equation was made only recently\cite{JSSM}-\cite{J2}. The formalism
of Jagannathan{\em et~al} was the first one to use the Dirac equation to
derive the focusing theory of electron lenses, in particular for magnetic
and electrostatic axially symmetric and quadrupole lenses respectively. 
This formalism further outlined the recipe to obtain a
quantum theory of aberrations. Details of these and some of the related
developments in {\em the quantum theory of charged-particle beam optics} can
be found in the references~\cite{JSSM}-\cite{Khan-QABP}. 
I shall briefly state the central theme of the quantum formalism.

The starting point to obtain a quantum prescription is to build a theory
based on the basic equations of quantum mechanics appropriate to the situation
under study. For situations when either there is no spin or spinor effects are 
believed to be small and ignorable we start with the scalar Klein-Gordon and
Schr\"{o}dinger equations for relativistic and nonrelativistic cases 
respectively. For electrons, protons and other spin-$\frac{1}{2}$ particles it
is natural to start with the Dirac equation, the equation for
spin-$\frac{1}{2}$ particles. In practice we do not have to care about the 
other~(higher spin) equations.

In many situations the electromagnetic fields are static or can reasonably
assumed to be static. In many such devices one can further ignore the times
of flights which are negligible or of not direct interest as the emphasis is
more on the profiles of the trajectories. The idea is to analyze the
evolution of the beam parameters of the various individual charged-particle
beam optical elements~(quadrupoles, bending magnets,~$\cdots$) along the optic
axis of the system. This in the language of the quantum formalism would 
require to know the evolution of the wavefunction of the beam particles as
a function of `$s$', the coordinate along the optic axis. Irrespective of the
starting basic time-dependent equation~(Schr\"{o}dinger, Klein-Gordon,
Dirac,~$\cdots$) the first step is to obtain an equation of the form
\bea
\i \hbar \frac{\partial }{\partial s} \psi \left(x\, , y\, ; s \right)
=
\hat{\cal H} \left(x\, , y ;\, s \right)
\psi \left(x\, , y\, ; s \right)\,,
\label{BOE}
\eea
where $\left(x , y ; s \right)$ constitute a curvilinear coordinate system,
adapted to the geometry of the system. For systems with straight optic axis,
as it is customary we shall choose the optic axis to lie along the $Z$-axis 
and consequently we have $s = z$ and
$\left(x , y ; z \right)$ constitutes a rectilinear coordinate system.
Eq.~(\ref{BOE}) is the basic equation in the quantum formalism and we call it
as the {\em beam-optical equation}; ${\cal H}$ and $\psi$ as the 
{\em beam-optical Hamiltonian} and the {\em beam wavefunction} respectively. The
second step requires to obtain a relationship for any relevant
observable $\left\{ \left\langle O \right\rangle \left( s \right) \right\}$ 
at the transverse-plane at $s$ to the observable 
$\left\{ \left\langle O \right\rangle \left( s_{\rm in} \right) \right\}$ 
at the transverse plane at $s _{\rm in}$, where $s _{\rm in}$ is some input
reference point. This is achieved by the integration of the beam-optical
equation in~(\ref{BOE})
\bea
\psi \left(x , y ; s \right) & = &
\hat{U} \left(s , s_{\rm in} \right)
\psi \left(x , y ; s_{\rm in} \right)\,, 
\label{BOI}
\eea
which gives the required transfer maps
\bea
\left\langle O \right\rangle \left(s_{\rm in} \right) 
& \rightarrow &
\left\langle O \right\rangle \left(s \right) \nn \\
& = &
\left\langle \psi \left(x , y ; s\right) 
\left| O \right|
\psi \left(x , y ; s \right)\right\rangle\,, \nn \\
& = &
\left\langle \psi \left(x , y ; s_{\rm in}\right) 
\left|\hat{U} ^{\dagger} O \hat{U}\right|
\psi \left(x , y ; s_{\rm in}\right)\right\rangle\,.
\label{BOM}
\eea

The two-step algorithm stated above may give an over-simplified picture of 
the quantum formalism than, it actually is. There are several crucial points 
to be noted. The first-step in the algorithm of obtaining the beam-optical 
equation is not to be treated as a mere transformation which eliminates~$t$ 
in preference to a variable~$s$ along the optic axis. A clever set of 
transforms are required which not only eliminate the variable $t$ in preference
to $s$ but also gives us the $s$-dependent equation which has a close physical
and mathematical analogy with the original $t$-dependent equation of standard
time-dependent quantum mechanics. The imposition of this stringent requirement 
on the construction of the beam-optical equation ensures the execution of the
second-step of the algorithm. The beam-optical equation is such, that all the
required rich machinery of quantum mechanics becomes applicable to compute the
transfer maps characterizing the optical system. This describes the essential
scheme of obtaining the quantum formalism. Rest is mostly a mathematical detail
which is built in the powerful algebraic machinery of the algorithm,
accompanied with some reasonable assumptions and approximations dictated by the
physical considerations. For instance, a straight optic axis is a reasonable 
assumption and paraxial approximation constitute a justifiable approximation to
describe the ideal behaviour. 

Before explicitly looking at the execution of the algorithm leading to the
quantum formalism in the spinor case, we further make note of certain other 
features. Step-one of the algorithm is achieved by a set of clever
transformations and an exact expression for the beam-optical Hamiltonian is 
obtained in the case of Schr\"{o}dinger, Klein-Gordon and Dirac equations
respectively, without resorting to any approximations! We expect this to be 
true even in the case of higher-spin equations. The approximations are made
only at step-two of the algorithm, while integrating the beam-optical
equation and computing the transfer maps for averages of the beam parameters.
Existence of approximations in the description of nonlinear behaviour is not 
uncommon and should come as no surprise, afterall the beam optics constitutes
a nonlinear system. The nature of these approximations can be best summarized 
in the optical terminology as; a systematic procedure of expanding the beam
optical Hamiltonian in a power series 
of $\left| {\hat{\vpi} _\perp}/{p_0} \right|$ where $p_0$ is the 
design~(or average) momentum of beam particles moving predominantly along the
direction of the optic axis and $\hat{\vpi} _\perp$ is the small transverse
kinetic momentum. The leading order approximation along with 
$\left| {\hat{\vpi} _\perp}/{p_0} \right| \ll 1$ constitutes the paraxial or 
ideal behaviour and higher order terms in the expansion give rise to the
nonlinear or aberrating behaviour. It is seen that the paraxial and
aberrating behaviour get modified by the quantum contributions which are in
powers of the de Broglie wavelength~($\loh = 2 \pi {\hbar}/{p_0}$). Lastly,
and importantly the question of the classical limit of the quantum formalism;
it reproduces the well known Lie algebraic formalism of 
charged-particle beam optics pioneered by Dragt~{\em et~al}~\cite{Lie}.

We start with the Dirac equation in the presence of static electromagnetic
field with potentials $\left(\phi (\r ), \A (\r ) \right)$
\beq
\hat{\rm H}_D \left| \psi_D \right\rangle = E 
\left| \psi_D \right\rangle\,,
\eeq
where $\left| \psi_D \right\rangle$ is the time-independent 
$4$-component Dirac spinor, $E$ is the energy of the beam particle 
and the Hamiltonian $\hat{\rm H}_D$, including the Pauli term in the 
usual notation is
\bea 
\hat{\rm H}_D & = & \beta m_0 c^2 
+ c \Al \cdot \hat{{\mbox{\boldmath $p$}}}
- \mu_a \beta \Vsig \cdot \B\,,
\eea
where $\hat{{\mbox{\boldmath $\pi$}}} = \hat{{\mbox{\boldmath $p$}}} - 
q{\mbox{\boldmath $A$}} = -i\hbar {\mbox{\boldmath $\nabla$}} - 
q{\mbox{\boldmath $A$}}$. After a series of 
transformations~(see~\cite{CJKP} for details) we obtain the accelerator 
optical Hamiltonian to the leading order approximation
\bea 
\i \hbar \ddz \left| \psi^{(A)} \right\rangle & = & \hH ^{(A)} 
\left| \psi^{(A)} \right\rangle\,, \nn \\ 
\hH ^{(A)} & \approx & \left( - p_0 - q A_z + \frac{1}{2 p_0}
\pitwo \right) \nn \\
& & 
+ \frac{\g m_0}{p_0} \Vomeg \cdot \S \,, \nn \\ 
{\rm with}\ \ \Vomeg & = & - \frac{1}{\g m_0} \left\{ q \B + 
\eps \left( \B _\parallel + \g \B _\perp \right) \right\}\,.
\eea 
where $\pitwo = \hat{\pi}_x^2 + \hat{\pi}_y^2$, 
$\eps = {2 m_0 \mu_a}/{\hbar}$, $\gamma = E/{m_0 c^2}$, and
$\S = \half \hbar \vsig$.  
We can recognize $\hH ^{(A)}$ as the {\em quantum mechanical, 
accelerator optical}, version of the well known semiclassical 
Derbenev-Kondratenko Hamiltonian~\cite{Heinemann} in the leading order
approximation. We can obtain corrections to this by going an order
beyond the first order calculation.  

It is straightforwrd to compute the transfer maps for a specific 
geometry and the detailed discussion with the quantum corrections
can be found in~\cite{CJKP}. In the classical limit we recover
the Lie algebraic formalism~\cite{Lie}.
 

One practical application of the quantum formalism would be to get a deeper
understanding of the polarized beams. A proposal to produce polarized beams
using the proposed spin-splitter devices based on the classical Stern-Gerlach
kicks has been presented recently~\cite{Splitter}.

Lastly it is speculated that the quantum theory of charged-particle beam
optics will be able to resolve the {\em choice of the position operator} 
in the Dirac theory and the related question of the {\em form of the force
experienced by a charged-particle in external electromagnetic 
fields}~\cite{Heinemann},~\cite{Barut}.
This will be possible provided one can do an extremely high precision 
experiment to detect the small differences arising in the transfer maps from
the different choices of the position operators. These differences shall be 
very small, i.e., proportional to powers of the de Broglie wavelength. It is
the extremely small magnitude of these minute differences which makes the
exercise so challenging and speculative!


\begin{thebibliography}{99}

\bibitem{HK}
P.W. Hawkes and E. Kasper,     
{\it Principles of Electron Optics}, Vols. I and II 
(Academic Press, London, 1989).

\bibitem{CM}
M. Conte and W.W. MacKay, 
{\it An Introduction to the Physics of Particle Accelerators}
(World Scientific, Singapore, 1991).

\bibitem{Mais} 
H. Mais, 
``Some topics in beam dynamics of storage rings'', 
DESY 96-119 (1996).
 
\bibitem{Wiedemann}
H. Wiedemann,
{\it Particle Accelerator Physics~: Basic Principles and
Linear Beam Dynamics}
(Springer-Verlag, Berlin, Heidelberg, 1993) \\
H. Wiedemann,
{\it Particle Accelerator Physics~II~: Nonlinear and
Higher-Order Beam Dynamics}
(Springer-Verlag, Berlin, Heidelberg, 1995)

\bibitem{HK3}
P.W. Hawkes and E. Kasper,
{\it Principles of Electron Optics}
Vol.3: {\it Wave Optics} (Academic Press, London and San Diego, 1994).

\bibitem{Chen}
P. Chen, {\it ICFA Beam Dynamics Newsletter} {\bf 12}, 46 (1996);

\bibitem{QABP}
{\em Proceedings of the 15th Advanced ICFA Beam Dynamics
Workshop on Quantum Aspects of beam Physics},
{\em Ed.} P.~Chen, (World Scientific, Singapore, 1999).

\bibitem{JSSM}
R.~Jagannathan, R.~Simon, E.~C.~G.~Sudarshan and N.~Mukunda,
{\em Phys. Lett. A} {\bf 134}, 457 (1989);
R. Jagannathan, in {\em Dirac and Feynman: Pioneers in  
Quantum Mechanics}, {\em Ed.} R.~Dutt and A.~K.~Ray (Wiley Eastern, New 
Delhi, 1993).

\bibitem{J2}
R. Jagannathan, 
{\em Phys. Rev. A} {\bf 42}, 6674 (1990).

\bibitem{KJ1}
S.A. Khan and R. Jagannathan, 
``Theory of relativistic electron beam transport based on the Dirac 
equation'', {\em Proc. of the 3rd National Seminar on Physics 
and Technology of Particle Accelerators and their Applications}, 
(Nov. 1993, Calcutta, India) Ed. S. N. Chintalapudi (IUC-DAEF, 
Calcutta) 102;
S.A. Khan and R. Jagannathan, 
``Quantum mechanics of charged-particle beam optics: An operator approach'',
Presented at the JSPS-KEK International Spring School on High Energy 
Ion Beams--Novel Beam Techniques and their Applications, 
March 1994, Japan, Preprint: IMSc/94/11 
(The Institute of Mathematical Sciences, Madras, March 1994). 

\bibitem{KJ3}
S.A. Khan and R. Jagannathan, 
{\em Phys. Rev. E} {\bf 51}, 2510 (1995).

\bibitem{JK2}
R. Jagannathan and S.A. Khan, 
{\em Advances in Imaging and Electron Physics}, {\bf 97}, 
Ed. P. W. Hawkes (Academic Press, San Diego) 257 (1996).    

\bibitem{Khan}
S.A. Khan,
{\it Quantum Theory of Charged-Particle Beam Optics}, Ph.D.
Thesis (University of Madras) (1997).

\bibitem{CJKP}
M. Conte, R. Jagannathan, S.A. Khan and M. Pusterla, 
{\em Part. Accel.} {\bf 56}, 99  (1996).

\bibitem{JK3}
R. Jagannathan and S.A. Khan 
{\em ICFA Beam Dynamics Newsletter} {\bf 13}, 21 (1997).

\bibitem{Jagan-QABP}
R.~Jagannathan,
``The Dirac equation approach to spin-$\frac{1}{2}$ particle beam
optics'', 
{\em in}: {\em Proceedings of the 15th Advanced ICFA Beam Dynamics
Workshop on Quantum Aspects of beam Physics},
{\em Ed.} P.~Chen, (World Scientific, Singapore, 1999),
physics/9803042.

\bibitem{Khan-QABP}
S.~A.~Khan,
{\bf Quantum theory of magnetic quadrupole lenses for 
spin-$\frac{1}{2}$ particles}, 
{\em in}: {\em Proceedings of the 15th Advanced ICFA Beam Dynamics
Workshop on Quantum Aspects of beam Physics},
{\em Ed.} P.~Chen, (World Scientific, Singapore, 1999),
physics/9809032.

\bibitem{Lie}
A.J. Dragt and E. Forest, 
{\em Adv. Electronics and Electron Phys.} {\bf 67}, 65 (1986);
A.J. Dragt, F. Neri, G. Rangarajan, D.R. Douglas, 
L.M. Healy and R.D. Ryne, 
{\em Ann. Rev. Nucl. Part. Sci.} {\bf 38}, 455 (1988);
G. Rangarajan, A.J. Dragt and F. Neri, 
{\em Part. Accel.} {\bf 28}, 119 (1990);
R.D. Ryne and A.J. Dragt,
{\em Part. Accel.} {\bf 35}, 129 (1991);
\'{E}. Forest and K. Hirata, 
{\em A Contemporary Guide to Beam Dynamics}
KEK Report 92-12;
\'{E}. Forest, M. Berz and J. Irwin, 
{\em Part. Accel.} {\bf 24}, 91 (1989);
K. Yokoya, ``Calculation of the equilibrium 
polarization of stored electron beams using Lie algebra'', 
Preprint KEK 86-90 (1986); 
Yu.I. Eidelman and V.Ye. Yakimenko,
{\em Part. Accel.} {\bf 45}, 17 (1994);
and references therein.


\bibitem{Heinemann}
``On Stern-Gerlach forces allowed by special relativity and the 
special case of the classical spinning particle of 
Derbenev-Kondratenko'', {\em e-print}: physics/9611001;
D.~P.~Barber, K.~Heinemann and G.~Ripken, {\it Z. Phys. C} {\bf 64} 
(1994) 117;
D.~P.~Barber, K.~Heinemann and G.~Ripken, {\it Z. Phys. C} {\bf 64} 
(1994) 143.

\bibitem{Splitter}
M.~Conte, A.~Penzo and M.~Pusterla, 
{\em Il Nuovo Cimento A} {\bf 108}, 127  (1995);
Y.~Onel, A.~Penzo and R.~Rossmanith, 
{\em AIP Conf. Proc.} 150
{\em Ed.} R.~G.~Lernerand and D.~F.~Geesaman,  
(AIP, New York, 1986) 1229;
M.~Conte, and M.~Pusterla, 
{\em Il Nuovo Cimento A}, {\bf 103}, 1087 (1990);
M.~Conte, Y.~Onel, A.~Penzo, A.~Pisent, M.~Pusterla and R.~Rossmanith,
The spin-splitter concept, Internal Report~: INFN/TC-93/04;
M.~Pusterla,
``Polarized beams and Stern-Gerlach forces in classical 
and quantum mechanics'', 
{\em in}: {\em Proceedings of the 15th Advanced ICFA Beam Dynamics
Workshop on Quantum Aspects of beam Physics},
{\em Ed.} P.~Chen, (World Scientific, Singapore, 1999),

\bibitem{Barut}
A.~O.~Barut and R.~Raczka,
{\it Theory of Group Representations and Applications}
(World Scientific, 1986);
J.~Anandan, {\em Nature} {\bf 387}, 558 (1997);
M.~Chaichian. R.~G.~Felipe and D.~L.~Martinez,  
{\em Phys. Lett.} A {\bf 236}, 188 (1997); 
J.~P.~Costella and B.~H.~J.~McKellar, 
{\em Int. J. Mod. Phys.} A {\bf 9}, 461 (1994); 
and references therein.  

\end{thebibliography}
\end{document}